\documentclass[12pt, dvips]{article}
\usepackage[english]{babel}
\usepackage{setspace}
\usepackage[dvips]{graphicx}
\usepackage[top=1in, bottom=1in, left=1in, right=1in]{geometry}
\usepackage{amsmath}
\usepackage[round, numbers, sort&compress]{natbib}
\usepackage{epsfig}

\begin{document}

\title{Hysteresis and bi-stability by an interplay of calcium oscillations and action potential firing}

\author{J.M.A.M. Kusters${^*}$, J.M. Cortes${^{*,\dag,\ddag}}$, W.P.M. van Meerwijk${^\S}$, D.L. Ypey${^\S}$,\\
 A.P.R. Theuvenet${^\S}$ and C.C.A.M. Gielen${^*}$}

\thispagestyle{empty}
\maketitle

\begin{tabular}{l}
\small{${^*}$ Dept. of Biophysics, Radboud University Nijmegen, Geert Grooteplein 21,}\\
\small{6525 EZ Nijmegen, The Netherlands}\\

\small{${^\dag}$ Institute Carlos I for Theoretical and
Computational
Physics}\\
\small{and Departamento de Electromagnetismo y Fisica de la
Materia.}\\
\small{Universidad de Granada, E-18071 Granada, Spain}\\

\small{${^\ddag}$ Institute for Adaptive and Neural Computation. School of Informatics,}\\
\small{University of Edinburgh, EH1 2QL, UK}\\

\small{${^\S}$ Department of Cell Biology, Radboud University Nijmegen, Toernooiveld 1,}\\
\small{6525 ED Nijmegen, The Netherlands}\\

\end{tabular}
\vfill
\hfill
\begin{tabular}{l}
Corresponding author: J.M.A.M. Kusters,\\
Dept. of Biophysics, Radboud University Nijmegen,\\
Geert Grooteplein 21, 6525 EZ Nijmegen, The Netherlands,\\
email: m.kusters@science.ru.nl, tel: +31-24-3615039, fax: +31-24-3541435\\
\end{tabular}
\pagebreak
\begin{abstract}

Many cell types exhibit oscillatory activity, such as repetitive
action potential firing due to the Hodgkin-Huxley dynamics of ion
channels in the cell membrane or reveal intracellular inositol
triphosphate (IP$_3$) mediated calcium oscillations (CaOs) by
calcium-induced calcium release channels (IP$_3$-receptor) in the
membrane of the endoplasmic reticulum (ER). The dynamics of the
excitable membrane and that of the IP$_3$-mediated CaOs have been
the subject of many studies. However, the interaction between the
excitable cell membrane and IP$_3$-mediated CaOs, which are coupled
by cytosolic calcium which affects the dynamics of both, has not
been studied. This study for the first time applied stability
analysis to investigate the dynamic behavior of a model, which
includes both an excitable membrane and an intracellular
IP$_3$-mediated calcium oscillator. Taking the IP$_3$ concentration
as a control parameter, the model exhibits a novel rich spectrum of
stable and unstable states with hysteresis. The four stable states
of the model correspond in detail to previously reported
growth-state dependent states of the membrane potential of normal
rat kidney fibroblasts in cell culture. The hysteresis is most
pronounced for experimentally observed parameter values of the
model, suggesting a functional importance of hysteresis. This study
shows that the four growth-dependent cell states may not reflect the
behavior of cells that have differentiated into different cell types
with different properties, but simply reflect four different
states of a single cell type, that is characterized by a single model.\\



\end{abstract}
\pagebreak

Complexity and multiple transitions among behaviorial states are
ubiquitous in biological systems \cite{murrayBOOK, keenerbook}. In
physics instabilities and hysteresis are well known to play an
important role in collective properties and have been studied since
many years \cite{haken1975,jones1976,velarde1977,hohenberg1993}.
Recently, multi-stability with hysteresis has also awakened a large
interest in biological systems \cite{ashwin2005}. Instabilities, for
instance, are crucial for efficient information processing in the
brain, such as in odor encoding \cite{huerta2001,laurent2001}.
Moreover, unstable dynamic attractors have been demonstrated in
cortical networks, with critical relevance to working memory and
attention \cite{torres2002,cortes2006,tsodyks2006}. In a wide sense,
multistable systems allow changes among different stable solutions
where the system takes advantage of instabilities as gateways to
switch between different stable branches \cite{ashwin2005}.
Bistability driven by instabilities prevents the system from
reaching intermediate states, e.g. partial mitosis. Hysteresis
prevents the system from changing its state when parameter values,
that characterize the system, vary. This is of relevance, for
instance, in cell mitosis. Once initiated, mitosis should not be
terminated before completion \cite{sha2003}. Thus, hysteresis may
lock the cell into a fixed state, preventing it from sliding back to
another state \cite{solomon2003}.

At the level of cell networks, multistability, and in particular
bistability, plays an important role in cell signaling as well
\cite{laurent1999, ferrell2004}. For example, communication between
neurons takes place at synaptic contacts, where arrival of an action
potential stimulates release of a neurotransmitter, thus affecting
the post-synaptic potential of the target cell. Typically, each cell
receives input from thousands of other cells mediated by different
neurotransmitters, which modify the post-synaptic potential by
excitation or inhibition at different time scales
\cite{ferrell2002}. This information at the cell membrane may be
transferred to the cell nucleus by so-called second messengers to
affect the nucleus in controlling DNA-expression, protein synthesis,
mitosis, etc. Calcium is one such second messenger and calcium
oscillations have been reported over a wide range of frequencies
with a chaotic or regular pattern \cite{chay1985}.

In many biological systems, cells display spontaneous calcium
oscillations (CaOs) and repetitive action-potential firing. These
phenomena have been described separately by models for intracellular
inositol trisphosphate (IP$_3$)-mediated CaOs \cite{deyoung1992,
sneyd2002} and for plasma membrane excitability \cite{torres2004}.
We have recently presented a single-cell model that combines an
excitable membrane with an IP$_3$-mediated intracellular calcium
oscillator \cite{kusters2005}. The IP$_3$-receptor is described as
an endoplasmic reticulum (ER) calcium channel with open and close
probabilities that depend on the cytoplasmic concentrations of
calcium ($[Ca_{cyt}^{2+}]$) and IP$_3$ ([IP$_3$]). An essential
component of this model relates to store-operated calcium channels
in the plasma membrane. Since it is not known whether multiple types
of  store-operated calcium channels are involved in normal rat
kidney (NRK) fibroblasts, we will use the general terminology of
store-dependent calcium (SDC) channels.

NRK fibroblasts in cell culture exhibit growth-state dependent
changes in their electrophysiological behavior \cite{harks2005}.
Subconfluent-grown serum-deprived quiescent cells exhibit a stable
resting membrane potential near -70 mV ("resting state"). Upon
subsequent treatment with epidermal growth factor the cells re-enter
the cell cycle, undergo density-dependent growth-arrest (contact
inhibition) at confluency and spontaneously fire action potentials
associated with intracellular calcium oscillations ("AP-firing
state"). Subsequent addition of retinoic acid or transforming growth
factor (TGF)$\beta$ to the contact inhibited cells causes the cells
to become phenotypically transformed and to depolarize the cell to
approximately -20 mV ("depolarized state"). This depolarization has
been shown to be caused by an elevation of the concentration of
prostaglandin (PG)F$_{2\alpha}$ secreted by the unrestricted
proliferating transformed cells. Washout of the medium conditioned
by the transformed cells by perfusion with fresh serum-free medium
causes the cells to slowly repolarize, and, preceded by a short
period of fast small-amplitude spiking of their membrane potential
("fast oscillating state"), to regain spontaneous repetitive action
potential firing activity ("AP-firing state") similar to that of the
contact inhibited cells. These phenomena have been described in
detail \cite{harks2005} and are very similar to the behavior of
other cell types with calcium oscillations and action potential
firing, such as interstitial cells of Cajal \cite{ward2000} and
hepatocytes \cite{dupont2003}.

\begin{figure}
\centerline{\psfig{file=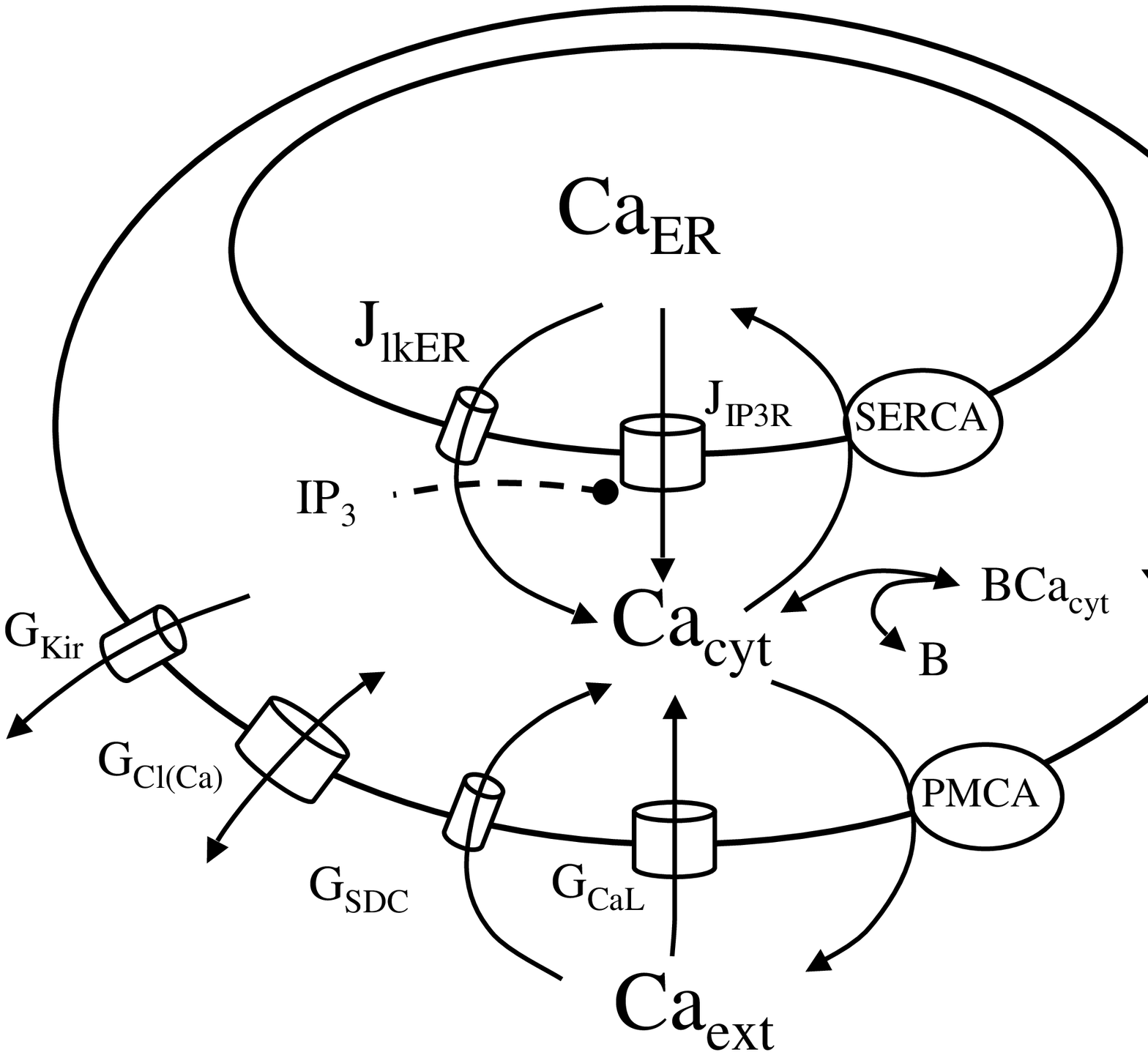,width=16cm}}\caption{}\label{fig1}
\end{figure}

In this study we have analyzed the model reported in
\cite{kusters2005}. This model, which is shown schematically in Fig.
\ref{fig1}, illustrates the basic characteristics of NRK
fibroblasts. It reproduces, on the basis of single-cell data
\cite{harks2003, kusters2005}, the dynamics of both the plasma
membrane excitability and that of the intracellular calcium
oscillator. We have recently shown that (PG)F$_{2\alpha}$
dose-dependently induces IP$_3$-dependent intracellular calcium
oscillations in NRK fibroblasts \cite{harks2003b}. Since the
growth-state dependent modulation of the membrane potential of NRK
fibroblasts is related to the concentration of (PG)F$_{2\alpha}$ in
their culture medium \cite{harks2005} and since this prostaglandin
dose-dependently increases [IP$_3$], we took [IP$_3$] as a control
parameter to analyze the stability of the single-cell model.

The stability analysis shows how coupling of an excitable membrane
with an intracellular calcium oscillator leads to a rich behavior of
a cell with multiple stable and unstable states with hysteresis. We
show that the growth-state dependent modulations of the membrane
potential of NRK fibroblasts in cell culture described above can be
understood as the stable states of the single-cell model with
membrane excitability and calcium oscillations of these cells. The
stable states of the model reproduce the four growth-dependent
states of NRK cells, corresponding to the resting state at $-70$ mV,
the AP-firing state for spontaneous action potential firing, the
depolarized state at $-20$ mV and the fast oscillating state with
small-amplitude spiking around $-20$ mV. Therefore, the four
growth-dependent states of NRK fibroblasts may not reflect the
behavior of cells that have differentiated into different cell types
with different properties, but reflect four different states of a
single cell type, that is characterized by a single model.

\section*{\small{Model description}}
The dynamics of NRK cell membrane excitability is given by a set of
equations which describe the active and passive ion transport
systems in the plasma membrane and the endoplasmic reticulum, as
illustrated in Fig. \ref{fig1} (see \cite{kusters2005,torres2004}
for a detailed description). The change in the membrane potential as
a function of time due to the currents through inwardly rectifying
potassium channels ($I_{Kir}$), L-type Ca-channels ($I_{CaL}$),
Ca-dependent Cl-channels ($I_{Cl(Ca)}$), leak channels $(I_{lk})$,
and SDC-channels ($I_{SDC}$) is given by
\begin{eqnarray}
C_m \frac{dV_m}{dt} = - (I_{Kir} + I_{lk} + I_{CaL}+I_{Cl(Ca)} +
I_{SDC}).
\end{eqnarray}
$I_{Kir}$ and $I_{lk}$ determine the membrane potential of the cell
at rest near -70 mV and are specified in \cite{kusters2005}.

The equation describing the L-type Ca-current $(I_{CaL})$ in terms
of the Hodgkin-Huxley kinetics of the L-type Ca-channel, is given by
\begin{eqnarray}\label{ICaL}
I_{CaL}= m \:h \:v_{Ca}\: G_{CaL} (V_m-E_{CaL}),
\end{eqnarray}
 where m is the voltage-dependent activation variable, h is the
voltage-dependent inactivation variable and $v_{Ca}$ is the
inactivation parameter. The dynamics of the variables m and h are
described by first order differential equations of the
Hodgkin-Huxley type \cite{kusters2005}. The calcium-dependent
inactivation is given by $v_{Ca} ~=~
K_{vCa}/([Ca_{cyt}^{2+}]+K_{vCa})$.

The Ca-dependent Cl-current $I_{Cl(Ca)}$ is given by
\begin{eqnarray}
I_{Cl(Ca)}=~ \frac{[Ca_{cyt}^{2+}]}{[Ca_{cyt}^{2+}] + K_{Cl(Ca)}}\:
G_{Cl(Ca)}\:(V_m -E_{Cl(Ca)})\label{dICldt}
\end{eqnarray}
The chloride current increases with the cytosolic calcium
concentration $[Ca_{cyt}^{2+}]$, causing a depolarization to the
Nernst potential of chloride ions ($E_{Cl(Ca)}$) near -20 mV in NRK
fibroblasts for sufficiently high values of $[Ca_{cyt}^{2+}]$.

The store-dependent calcium current $I_{SDC}$ is described by
\begin{eqnarray}
I_{SDC}=\: \frac{K_{SDC}}{[Ca_{ER}^{2+}] + K_{SDC}} \: G_{SDC}\:(V_m
- E_{SDC})\label{ISDC}.
\end{eqnarray}
This store-dependent calcium channel allows calcium ions to flow
from the extracellular space into the cytosol at a rate inversely
proportional to the calcium concentration in the ER
\cite{hofer1998}. SDC channels are thought to play a major role in
the control of Ca-homeostasis in
the cell \cite{kusters2005, feske2006}.\\

The rate of change of Ca-content of the cytosol of the cell due to
inflow through the cell membrane and from the ER store, and by
buffering is described by
\begin{eqnarray}Vol_{cyt} \frac{d[Ca_{cyt}^{2+}]}{dt}   &=& A_{PM}\:J_{PM}+A_{ER} (
J_{IP_3R}+J_{lkER} - J_{SERCA}) - Vol_{cyt}\:
\frac{d[BCa]}{dt},\label{dCacytdt}
\end{eqnarray} where $Vol_{cyt}$ represents the cytoplasmic volume and
$A_{PM}$ and $A_{ER}$ the area of the cell membrane and of the ER
membrane, respectively. The term [BCa] denotes the buffer-calcium
complex in the cytosol and will be explained later. The flux of
calcium through the membrane ($J_{PM}$) is the sum of the influxes
of $Ca^{2+}$ ions through the L-type Ca-channel, through the
SDC-channel, and of the extrusion by the PMCA-pump
\cite{kusters2005}, and is given by $ J_{PM} = - (1/(z_{Ca} F
A_{PM})) (I_{CaL} + I_{SDC})- J_{PMCA}$.

The dynamics for the intracellular calcium oscillator is described
by the flux of calcium through the ER membrane. The rate of change
of calcium content in the ER depends on the sum of flux through the
IP$_3$-receptor ($J_{IP_3R}$), flux by leak through the ER-membrane
($J_{lkER}$) and flux by removal by the SERCA pump ($J_{SERCA}$),
which results in
\begin{eqnarray}
Vol_{ER}\frac{d[Ca_{ER}^{2+}]}{dt} = A_{ER} (- J_{IP_3R} - J_{lkER}
+ J_{SERCA}),
\end{eqnarray}
where $Vol_{ER}$ represents the volume of the ER.

The flux through the IP$_3$-receptor is described by
\begin{eqnarray}
J_{IP_3R} =~ f_{\infty}^3\: w ^3 \: K_{IP_3R} \: ([Ca_{ER}^{2+}]
-[Ca_{cyt}^{2+}])\label{JIP3R}
\end{eqnarray}
where $[Ca_{ER}^{2+}] -[Ca_{cyt}^{2+}]$ is the concentration
difference between calcium in the ER and in the cytosol. $K_{IP_3R}$
is the rate constant per unit area of IP$_3$-receptor mediated
release. The terms $f_{\infty}$ and $w$ represent the fraction of
open activation and inactivation gates, respectively. $f_{\infty}$
and $w_{\infty}$ depend both on the cytosolic calcium concentration
and are described by
\begin{eqnarray}
f_{\infty}         ~=~ \frac{[Ca_{cyt}^{2+}]}{K_{fIP_3}
+[Ca_{cyt}^{2+}]}
\end{eqnarray}
and
\begin{eqnarray}
w_{\infty} ~=~ \frac{\frac{[IP_3]}{K_{wIP_3} + [IP_3]}}{
\frac{[IP_3]}{K_{wIP_3}
+[IP_3]}+K_{w(Ca)}[Ca_{cyt}^{2+}]}\label{winfty}.
\end{eqnarray}
The inactivation time constant of the IP$_3$-receptor is defined by
\begin{eqnarray}
\tau_w   ~=~ \frac{a}{\frac{[IP_3]}{K_{wIP_3} +
[IP_3]}+K_{w(Ca)}[Ca_{cyt}^{2+}]}.\label{tauw}
\end{eqnarray}
$K_{fIP_3}$, $K_{wIP_3}$, $K_{w(Ca)}$ and $a$ are constants. The
fraction of open activation gates (f) is independent of the IP$_3$
concentration, but increases when the calcium concentration in the
cytosol increases. The fraction of open inactivation gates (w)
depends on the IP$_3$ concentration and on [$Ca_{cyt}^{2+}$].
$\tau_w$ determines the duration of the de-inactivation of w.

 $J_{lkER}$ is a passive leak of $Ca^{2+}$ from the ER into the cytosol which is not
mediated by the IP$_3$-receptor, but by an additional Ca-channel in
the ER membrane, presumably the translocon. Experimental evidence
for a role of the translocon complex as a passive $Ca^{2+}$ leak
channel has been presented recently \cite{flourakis2006}. $J_{lkER}$
is given by $J_{lkER} \:=\: K_{lkER} ([Ca_{ER}^{2+}] -
[Ca_{cyt}^{2+}])$. We used the leakage parameter $K_{lkER}$ as a
control parameter to study the dynamics of the plasma membrane,
because changes in the leak of Ca-ions through the ER membrane
produce proportional changes in [$Ca_{cyt}^{2+}$].

$J_{SERCA}$ represents the flux of calcium into the ER by the SERCA
pump and is given by $J_{SERCA}=~ J_{SERCA}^{max} \:
\{[Ca_{cyt}^{2+}]^2 /
(K_{SERCA}^2+[Ca_{cyt}^{2+}]^2)\}\label{JSERCA}$.

Finally, calcium in the cytosol is buffered by proteins in the
cytosol. The dynamics of buffering is given by $d[BCa]/dt \:=\:
k_{on}([T_B] - [BCa])[Ca_{cyt}^{2+}] -k_{off}\:[BCa]\label{buffer}$,
where $[T_B]$ is the total concentration of buffer in the cytosol
and $k_{on}$ and $k_{off}$ are the buffer rates \cite{kusters2005}.

The excitable membrane and the IP$_3$-mediated intracellular calcium
oscillator are coupled by the Ca-concentration $[Ca_{cyt}^{2+}]$ in
the cytosol as explained in \cite{kusters2005}. During an action
potential, opening of the L-type Ca-channel causes a large inward
current of Ca-ions through the plasma membrane. The increased
$[Ca_{cyt}^{2+}]$ activates the IP$_3$-receptor (calcium release
channel), causing calcium release from the ER, which further
contributes to the intracellular cytosolic calcium transient. In the
reverse process, IP$_3$-mediated calcium oscillations cause periodic
calcium transients, which lead to periodic opening of the
Ca-dependent Cl-channels. The depolarization of the membrane
potential towards the Nernst potential of the Ca-dependent
Cl-channels near -20 mV causes activation of the L-type Ca-channels
in the plasma membrane and excitation \cite{torres2004,
deroos1997b}. After an action potential or Ca-transient the
reduction of cytosolic calcium by the activity of the SERCA and PMCA
pumps reduces $I_{Cl(Ca)}$ (see Eq. \ref{dICldt}), enough to allow
the membrane to return to the membrane potential at rest near $-70$
mV.

The dynamics of the single-cell model depends on seven variables
($m$, $h$, $w$, $[BCa]$, $V_m$, $[Ca_{cyt}^{2+}]$ and
$[Ca_{ER}^{2+}]$), which were defined above. To study the stability
of the complete system we have determined the singular states for
the system and calculated the Floquet multipliers of these singular
states \cite{fairgrave1991,ioossBOOK}.

\section*{\small{Stability analysis of the membrane model}}
We will first analyze the bifurcations and local stability of both
the excitable membrane and intracellular calcium oscillator
separately, and then compare the results with the properties of the
single-cell model including both the membrane dynamics and
intracellular calcium oscillator. Two different analyses, namely,
our own implementation in $C$, and the software package $XPPAUT$
\cite{ermentroutBOOK}, which includes an $AUTO86$ \cite{doedelAUTO}
interface, gave the same results.

\begin{figure}
\centerline{\psfig{file=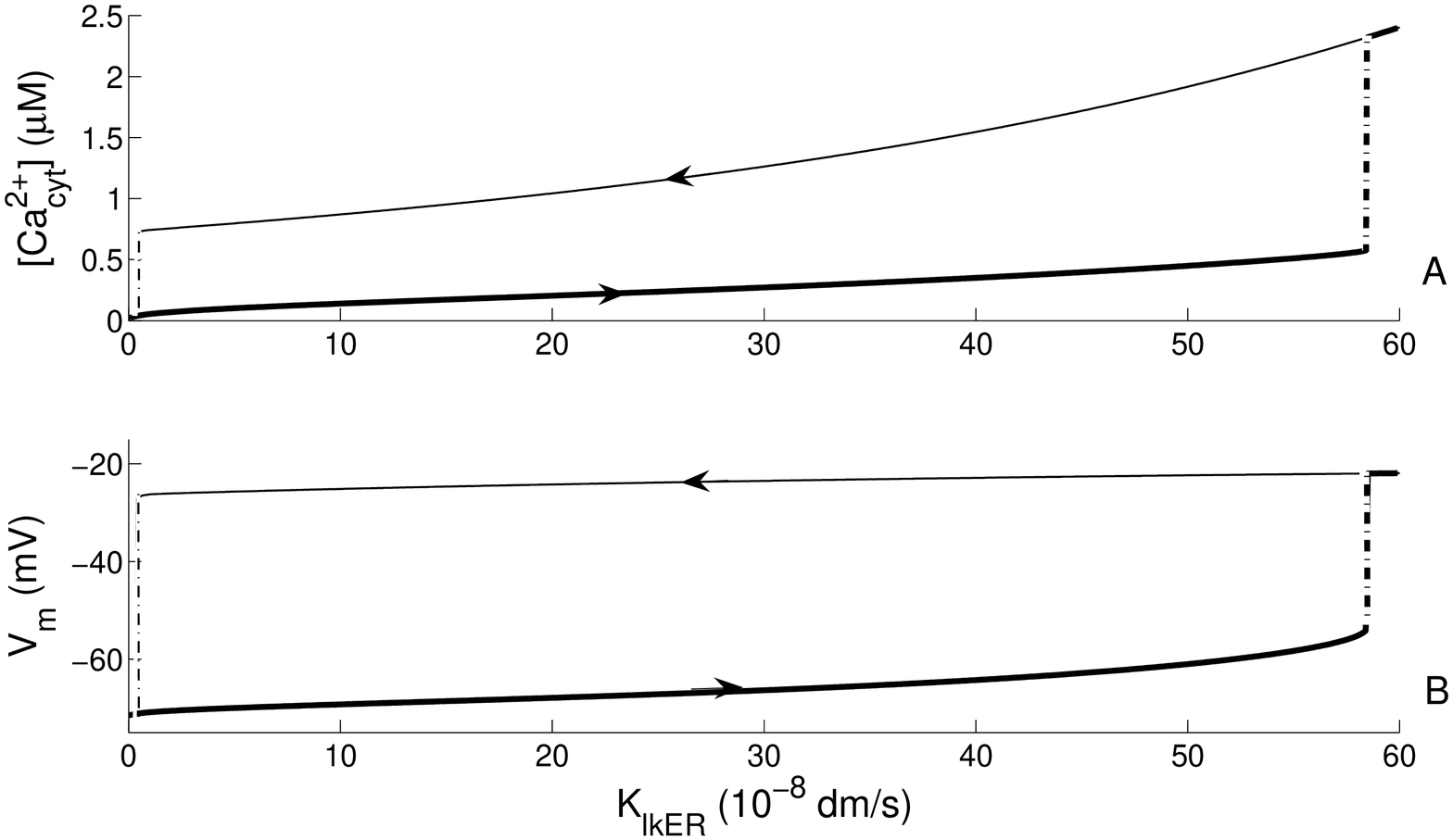,width=16cm}}\caption{}\label{fig2}
\end{figure}

In the single-cell model the intracellular calcium oscillations can
be eliminated by setting the IP$_3$ concentration ([IP$_3$]) to
zero. This allows the study of the excitable cell membrane
separately from the calcium oscillator. The dynamics of the plasma
membrane depends on the cytosolic calcium concentration
[$Ca_{cyt}^{2+}$], which opens the Ca-dependent Cl-channel. Since
the leak of Ca-ions from the ER affects the mean value of
[$Ca_{cyt}^{2+}$], the dynamics of the membrane is studied as a
function of the leakage parameter $K_{lkER}$. Fig. \ref{fig2} shows
a hysteresis diagram for the excitable cell membrane with the steady
states of the calcium concentration in the cytosol
([$Ca_{cyt}^{2+}$], panel A) and of the membrane potential ($V_m$,
panel B). The thick and thin solid lines refer to the stable states
for increasing and decreasing values of $K_{lkER}$, respectively.
The dashed-dotted lines reflect the transitions between the two
stable branches for increasing and decreasing values of $K_{lkER}$.

Starting at the value zero for $K_{lkER}$, the inwardly rectifying
K-channels keep the membrane potential at the resting membrane
potential of the NRK fibroblasts near $-70$ mV, where the membrane
is able to produce an action potential upon electrical stimulation
\cite{kusters2005}. For increasing values of $K_{lkER}$,
[$Ca_{cyt}^{2+}$] and $V_m$ increase gradually, causing a decreasing
threshold for activation. The gradual increase of $V_m$ is due to
gradual opening of the Ca-dependent Cl-channels for increasing
[$Ca_{cyt}^{2+}$] (see Eq. \ref{dICldt}). At $K_{lkER} \approx 58.0
\times 10^{-8}$ $dm/s$, [$Ca_{cyt}^{2+}$] is large enough to open
the Ca-dependent Cl-channels driving the membrane potential towards
the Nernst potential for Cl$^{-}$-ions which is near $-20$ $mV$ in
NRK fibroblasts (see Fig. \ref{fig2}B). The resulting depolarization
causes closure of the inwardly rectifying K-channels and opening of
the L-type Ca-channels which leads to an increase of calcium inflow
from the extracellular medium into the cytosol. The positive
feedback via the membrane potential between Ca-dependent Cl-channels
and L-type Ca-channels explains the abrupt increase of
[$Ca_{cyt}^{2+}$] (dashed-dotted line) to $2.3$ $\mu M$.

When we decrease $K_{lkER}$ starting from $60.0 \times 10^{-8}$
$dm/s$ (thin solid line), the cell remains depolarized near -20 mV
far below the value of $K_{lkER}$ at $58.0 \times 10^{-8}$ $dm/s$.
This is caused by the feedback between the Ca-dependent Cl-channels
and L-type Ca-channels. When $K_{lkER}$ decreases, [$Ca_{cyt}^{2+}$]
also decreases, which reduces the fraction of open Ca-dependent
Cl-channels. As a consequence, the membrane potential slightly
decreases just below -20 mV, which leads to an increased fraction of
open L-type Ca-channels, since the product of mh of steady-state
activation and inactivation (see Eq. \ref{ICaL}) reaches a maximum
just below -20 mV. The increment of the fraction of open L-type
Ca-channels leads to an extra inflow of calcium in the cytosol,
which increases the fraction of the open Ca-dependent Cl-channels
and prevents the system from falling back to a membrane potential
near $-70$ mV. Thus, in spite of the slow decrease of calcium
concentration and membrane potential caused by the leak channels,
the feedback by the L-type Ca-channels keeps the system at an
elevated [$Ca_{cyt}^{2+}$] and membrane potential near $-20$ mV
until low values $K_{lkER}$. The calcium in the cytosol returns to a
low concentration, only when the $K_{lkER}$ is decreased to very low
values. Then the Ca-dependent Cl-channels close and the membrane
potential repolarizes to $-70$ $mV$. Separate simulations showed
that the inward rectifier contributes to the transitions, but not to
the hysteresis.

\section*{\small{Stability analysis of the intracellular calcium oscillator}}
Following a similar plan as for the excitable cell membrane, we
obtained a bifurcation diagram for the intracellular calcium
oscillator as a function of the IP$_3$ concentration under
conditions that the L-type Ca-channels are blocked such as with
nifedipine. This was achieved by setting G$_{CaL}$ to zero and
K$_{lkER}$ to its physiological value of $2.0 \times 10^{-8}$
$dm/s$. In this way, we eliminate the contribution of calcium inflow
by the L-type Ca-channels and remove a principal influence of the
membrane model on the intracellular calcium oscillator. Therefore,
we only take into account the Ca-flux through the SDC-channels and
PMCA pump in the plasma membrane.

\begin{figure}
\centerline{\psfig{file=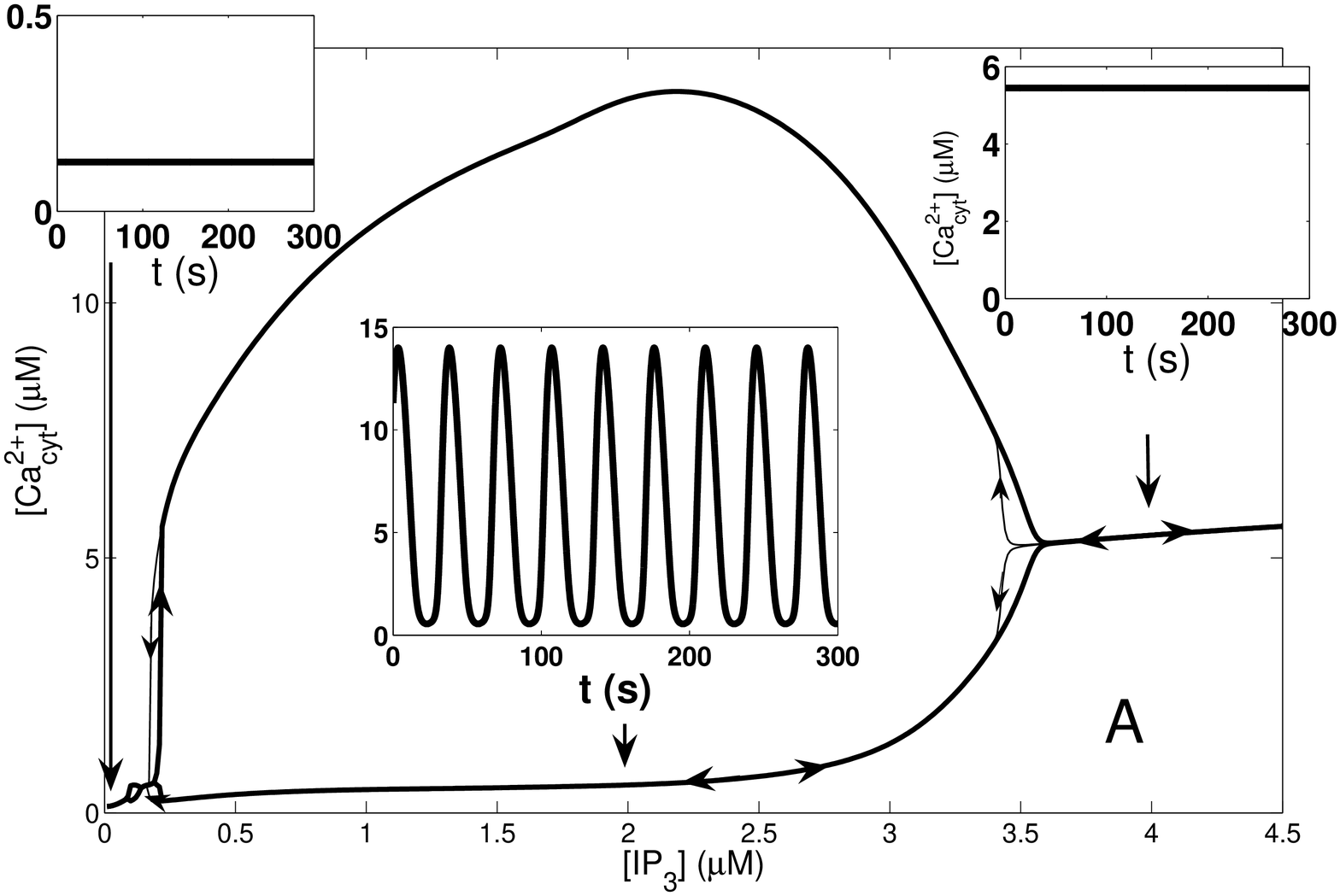,width=14cm}}
\centerline{\psfig{file=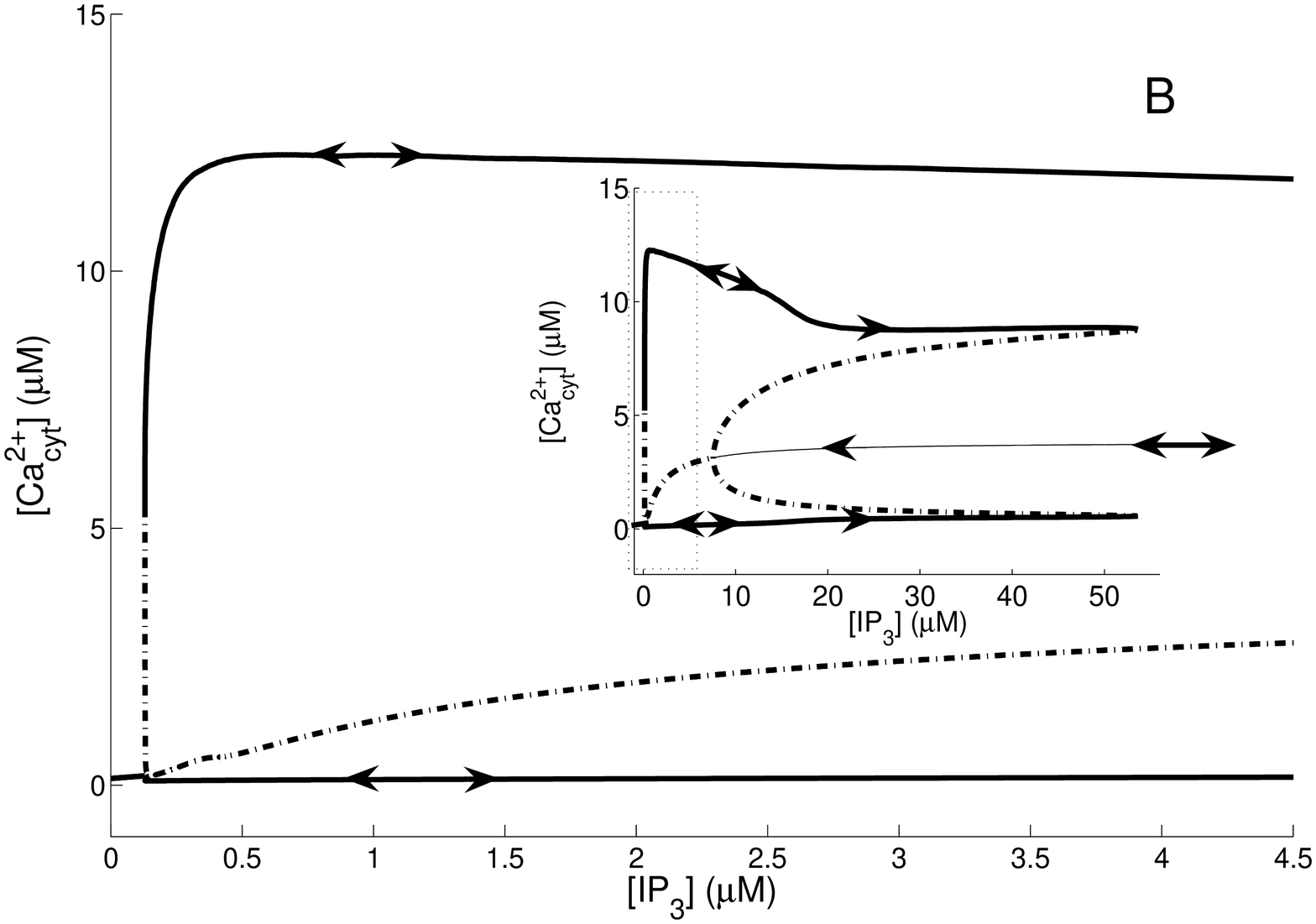,width=14cm}}
\caption{}\label{fig3}
\end{figure}

As explained in \cite{kusters2005}, the relative strength of the
PMCA and SERCA pump is crucial to reproduce the steady state calcium
concentrations in the cytosol and in the ER. By eliminating the
calcium inflow by the L-type Ca-channels, less calcium flows into
the cell. Therefore, we have to change the relative strength of the
PMCA and/or SERCA pump to maintain the proper balance between
calcium concentration in the cytosol and ER. In this model study, we
choose to decrease the strength of the SERCA pump. By doing so, the
system reveals a bifurcation diagram (Fig. \ref{fig3}A) similar to
that observed in other models \cite{li2002,schuster2002}.

Fig. \ref{fig3}A shows the dynamical behavior of [$Ca_{cyt}^{2+}$]
as a function of [IP$_3$] for $G_{CaL} = 0$ and with $J_{SERCA}^{
max}$ set to 2 x $10^{-5}$ $(\mu mol)/(s$ x $dm^2)$. Fig.
\ref{fig3}A shows a single stable steady state for small values of
[IP$_3$] (range 0.0 - 0.2 $\mu M$). At [IP$_3$] near 0.2 $\mu M$ the
dynamics reveals a supercritical Hopf bifurcation (thick solid
line), and the system becomes a calcium oscillator in the range of
IP$_3$ concentrations between 0.2 and 3.6 $\mu M$. For [IP$_3$]
values near $3.6$ $\mu M$ the system meets a supercritical Hopf
bifurcation and remains stable for higher IP$_3$ concentration at a
Ca-concentration near $5$ $\mu M$. In the range for [IP$_3$] above
3.5 $\mu M$, the elevated mean level of [$Ca_{cyt}^{2+}$] gives rise
to a short time constant $\tau_w$ for the inactivation parameter $w$
(Eq. \ref{tauw}). Due to this small time constant the inactivation
$w$ recovers relatively fast compared to the removal of
[$Ca_{cyt}^{2+}$], i.e. before the activation parameter $f$
de-activates to small values. As a result the product $f w$ does not
reach small values and the IP$_3$-receptor remains open (see Eq.
\ref{JIP3R}), causing a constant leak of calcium.

For decreasing [IP$_3$] values (thin solid line), the system starts
at a stable fixed point which remains stable until $3.50$ $\mu M$.
In the range [IP$_3$] between 3.5 and 3.6 $\mu M$, the system
exhibits bistability and a hysteresis over a small range of
IP$_3$-values. This hysteresis is caused by the positive feedback
between [$Ca_{cyt}^{2+}$] and the activation gate ($f$). For
decreasing [IP$_3$], the [$Ca_{cyt}^{2+}$] is already elevated and
so a large fraction of activation gates $f$ is already open and the
time constant $\tau_w$ is short. Due to the short time constant
$\tau_w$, the time for de-inactivation (w) is faster than for
de-activation (f). As a consequence, the product of $f$ and $w$ does
not reach small values and calcium passes continuously through the
IP$_3$-receptor from the store into the cytosol. This hysteresis did
not show up in the figures presented by Li $\&$ Rinzel
\cite{li2002}, but appears in their model if we insert the parameter
values which apply to the NRK fibroblasts (see \cite{kusters2005}).

For [IP$_3$] values below 3.5 $\mu M$, the stable fixed point
disappears, and the system starts to operate as an oscillator, until
[IP$_3$] values smaller than $0.15$ $\mu M$, where the system
returns to a single stable steady state.

As a next step, we have set the strength of the SERCA pump back to
its default value 8.$10^{-5}$ $\mu mol/(s$ x $dm^2)$ which
corresponds to the value in the single-cell model with an excitable
membrane and IP$_3$-mediated calcium oscillations. This results in
the bifurcation diagram shown in Fig. \ref{fig3}B. Fig. \ref{fig3}B
shows a major hysteresis in the [IP$_3$] range between 8 and 53 $\mu
M$ (see inset). To compare the results with those in Fig.
\ref{fig3}A we scaled Fig. \ref{fig3}B in the same [IP$_3$] range as
in Fig. \ref{fig3}A.

When the strength of the SERCA pump is increased to 8.$10^{-5}$ $\mu
mol/(s$ x $dm^2)$, [$Ca_{cyt}^{2+}$] decreases more rapidly after a
calcium transient. This affects the time constant $\tau_w$ of the
inactivation parameter $w$ (see Eq. \ref{tauw}). For small
[$Ca_{cyt}^{2+}$] levels, this time constant is relatively large,
ensuring a slow de-inactivation. This explains why a more powerful
SERCA pump gives rise to calcium oscillations over a much larger
range of IP$_3$ concentrations. Only at sufficiently large [IP$_3$]
values does $\tau_w$ become sufficiently small such that
de-inactivation ($w$) takes place more rapidly than de-activation
($f$). For these high IP$_3$-values, the product fw of the
activation parameter (f) and the inactivation parameter (w) is large
enough to allow a continuous leak of calcium through the
IP$_3$-receptor.

The inset in Fig. \ref{fig3}B shows a single stable steady state for
small values of [IP$_3$]. At [IP$_3$] near 0.2 $\mu M$, the dynamics
reveals a subcritical Hopf bifurcation (thick solid line), and the
system becomes a calcium oscillator in the range of IP$_3$
concentrations between 0.2 and 53 $\mu M$. For [IP$_3$] above 53
$\mu M$, [$Ca_{cyt}^{2+}$] is elevated at a steady state
concentration near 4 $\mu$M. For decreasing [IP$_3$] values (thin
solid line), the system starts at a stable elevated
[$Ca_{cyt}^{2+}$] which remains stable until [IP$_3$] is near 8 $\mu
M$.
For [IP$_3$] values below 8 $\mu M$, the stable fixed point
disappears, and the system starts to operate as an oscillator, until
[IP$_3$] values smaller than 0.2 $\mu M$. We conclude that
increasing the activity of the SERCA pump makes it more easy for the
cell to oscillate at higher [IP$_3$] values and causes a hysteresis
over a larger range of [IP$_3$] values.




\section*{\small{Stability analysis of the single-cell model}}

\begin{figure}
\centerline{
\psfig{file=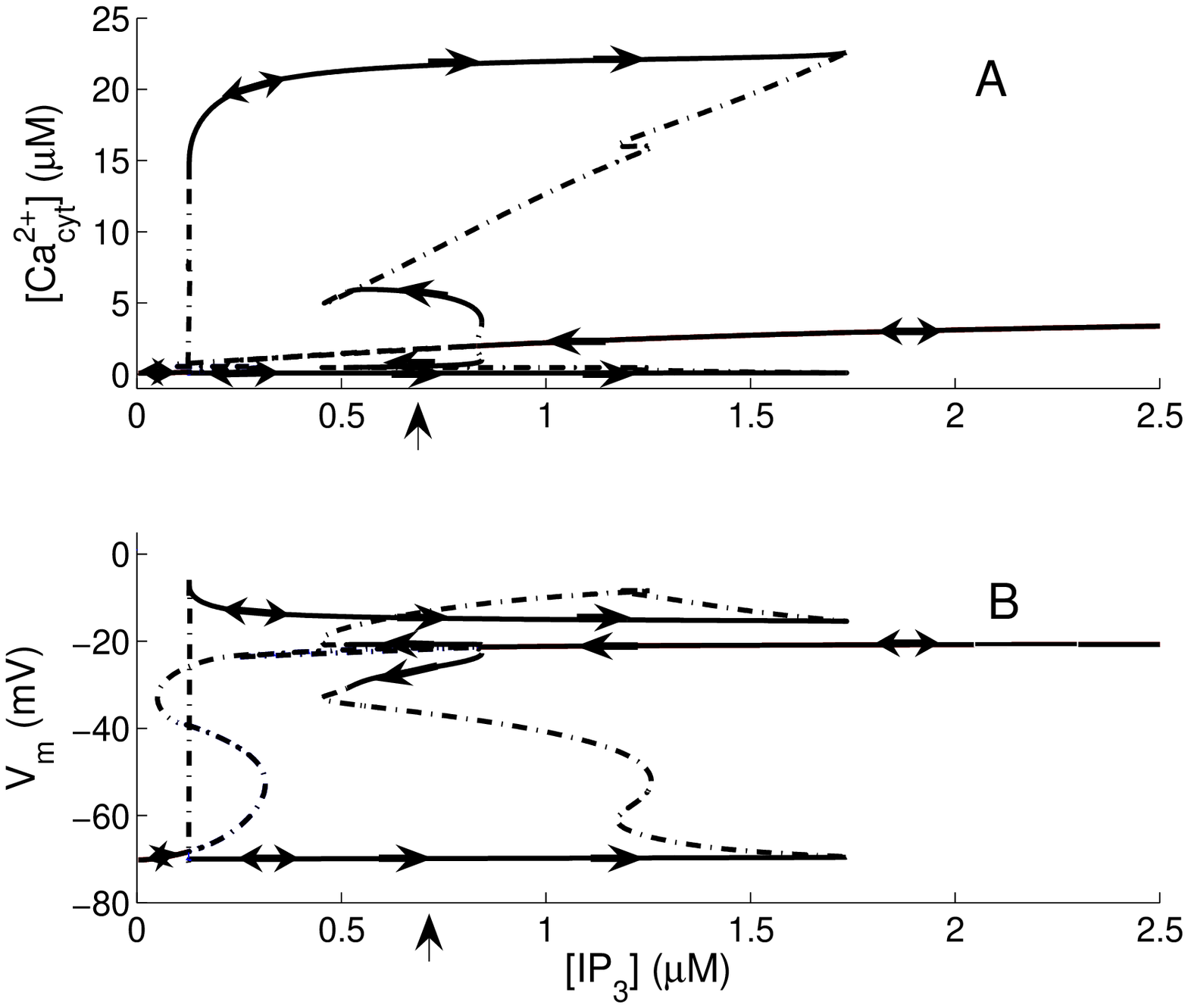,width=10.5cm}}
\centerline{\psfig{file=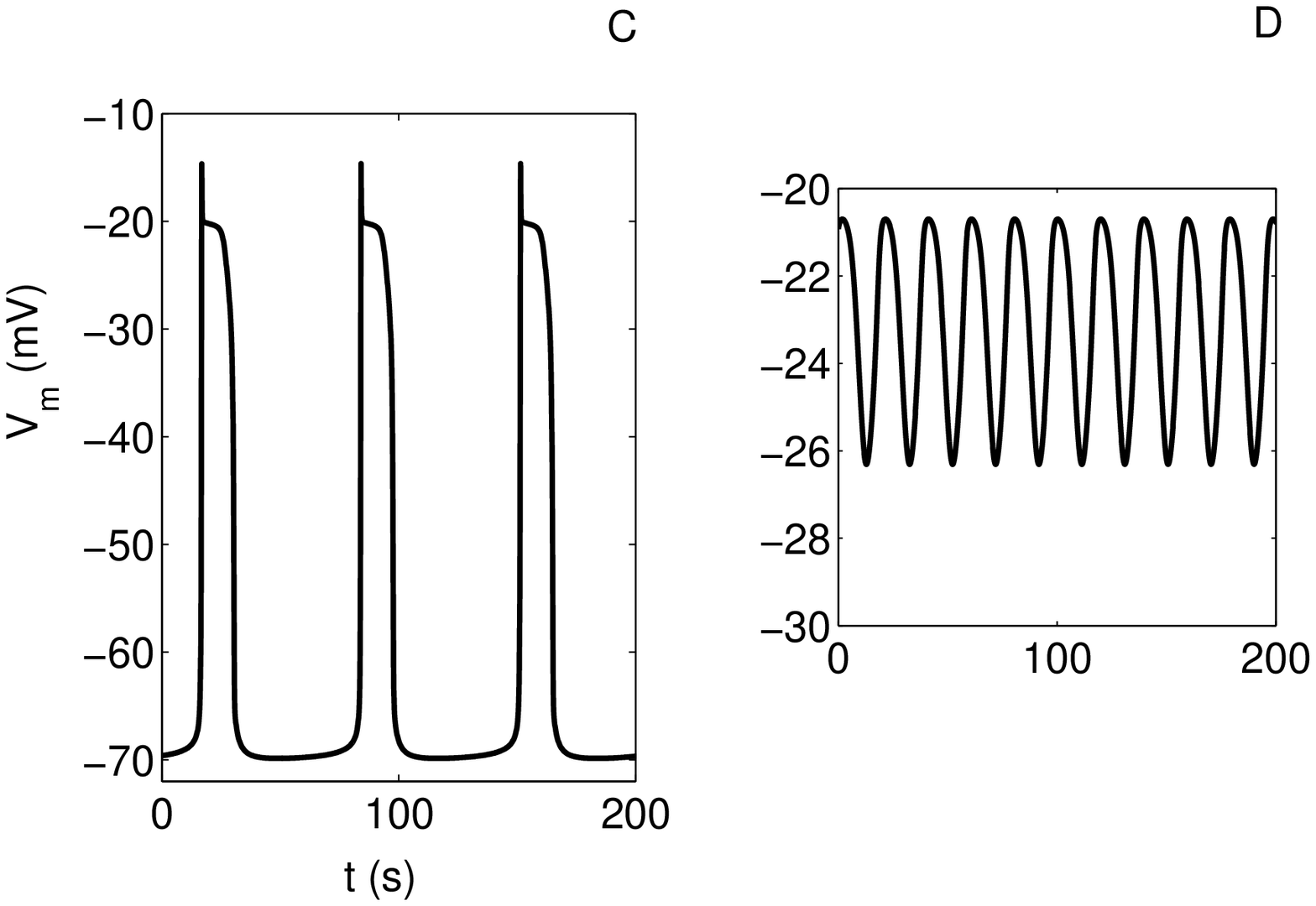,height=8cm,width=10.5cm}
}\caption{} \label{fig4}
\end{figure}

Unblocking the L-type Ca-channels ($G_{CaL}=$0.7 $\mu M$) transforms
the bifurcation diagram of \ref{fig3}B into that of Fig.
\ref{fig4}A. Fig. \ref{fig4} shows [$Ca_{cyt}^{2+}$] (panel $A$) and
the membrane potential (panel $B$) as a function of IP$_3$
concentration in the cell. The solid and dashed-dotted lines
represent stable and unstable states, respectively. For small
[IP$_3$] values in the range from 0.00 to 0.15 $\mu M$, the cell has
a single stable steady state ("resting state") with a membrane
potential near {-70} mV. For [$IP_3]
> 0.15$ $\mu M$, the stable fixed point becomes unstable in a
subcritical Hopf bifurcation. Calcium oscillations together with
action potentials occur for IP$_3$ concentrations in the range
between 0.15 and 1.75 $\mu M$ ("AP-firing state") (see panel $C$
which shows the membrane potential as a function of time for
[IP$_3$] $=$ 0.7 $\mu$M). In this regime, a rapid calcium inflow
from the $ER$ into the cytosol opens the Ca-dependent Cl-channel,
causing an inward current towards the Cl-Nernst potential close to
$-20$ $mV$. This depolarization activates the L-type Ca-channels
leading to an AP. After closure of the IP$_3$-receptor, calcium is
removed from the cytosol by the Ca-pumps in the cell membrane and
ER, leading to repolarization to $-70$ mV. For [IP$_3$] $>$ 1.75
$\mu M$, the fixed point $([Ca_{cyt}^{2+}],V_m)$ near (3.00 $\mu
M$,-20 $mV)$ becomes stable in a subcritical Hopf bifurcation
("depolarized state"). This can be understood from the fact that the
time-constant $\tau_w$ (see Eq. \ref{tauw}) for calcium-dependent
(de-)inactivation of the IP$_3$-receptor decreases for increasing
values of [IP$_3$] and for increasing values of the mean
$[Ca_{cyt}^{2+}]$. Near [IP$_3$] $=$ 1.75, the time-constant
$\tau_w$ is relatively short. During a cytosolic calcium transient
the fast inactivation of the inactivation gates $w$ of the
IP$_3$-receptor is followed by a fast de-inactivation of the
inactivation gates of the IP$_3$-receptor. During the fast
de-inactivation, the fraction of open activation gates $f$ of the
IP$_3$-receptor is still high due to high $[Ca_{cyt}^{2+}]$ (because
removal of calcium through the SERCA and PMCA pump is not fast
enough). As a consequence the IP$_3$-receptor remains open. Now the
$IP_3$-receptor acts as a constant leak channel, like $J_{lkER}$.
This leak of calcium into the cytosol opens the Ca-dependent
Cl-channels, causing a maintained depolarization to the Cl-Nernst
potential near $-20$ $mV$ ("depolarized state") (panel B).

If [IP$_3$] is decreased starting from [IP$_3$] $=$ 2.5 $\mu M$, the
cell with both the excitable membrane and intracellular calcium
oscillator active exhibits a complex hysteresis pattern. For
decreasing IP$_3$ concentrations, the system stays in a single
stable state (solid line) at an elevated $[Ca_{cyt}^{2+}]$ near 3
$\mu M$ and a membrane potential near $-20$ $mV$ until [IP$_3$]
$\approx$ 0.85 $\mu M$ ("depolarized state"). Then, the cell goes
through a Hopf bifurcation (dashed line) forcing the system to
behave as a stable oscillator with small calcium oscillations with
an amplitude of about $6$ $\mu$M and with small membrane potential
oscillations around $- 23$ mV ("fast oscillating state"). These
small oscillations of the membrane potential just below $-20$ mV are
illustrated in more detail in panel D. Note that the oscillations of
the membrane potential in panels C and D are both obtained for
[IP$_3$] $=$ 0.7 $\mu$M, illustrating the hysteresis. The
oscillations shown in panel D are due to small IP$_3$-mediated
calcium oscillations with active involvement of the L-type
Ca-channel dynamics. Due to decreasing [IP$_3$], the de-inactivation
time constant $\tau_w$ of the IP$_3$-receptor increases gradually.
This makes it possible for the cell to generate calcium
oscillations. Setting $v_{ca}$ in the equation for the L-type
Ca-current to 1 does not change the bifurcation diagram of Fig.
\ref{fig4}. The shape of the bifurcation scheme in Fig. \ref{fig4}A
remains the same, but the calcium oscillations extend over a larger
range of $[Ca_{cyt}^{2+}]$-values (approximately twice as large).

At [IP$_3$] $\approx$ 0.45 $\mu M$ the stable small-amplitude
oscillator becomes unstable (dashed line), returning the system to
the stable oscillations with large amplitude Ca-oscillations with a
peak value near $20$ $\mu M$ and with action potentials in the range
between -70 and -10 $mV$ ("AP-firing state"). Finally, for [IP$_3$]
values smaller than $0.15$ $\mu M$ the system returns to a single
stable state ("resting state").

In comparison with the simple dynamics of the cell membrane and
intracellular Ca-oscillator, shown in Figs. \ref{fig2} and
\ref{fig3}, it is remarkable to see the complex behavior of the
single-cell model shown in Fig. \ref{fig4}.

Since the SDC channels in the plasma membrane play a crucial role in
stabilization of the calcium dynamics \cite{kusters2005,
mignen2005}, we studied the dynamics of the cell as a function of
the SDC conductance in a range between 0.00 and 0.20 nS. Fig.
\ref{fig5} shows the hysteresis diagrams for five different values
of $G_{SDC}$. As explained in \citep{kusters2005}, the calcium
homeostasis of the cell is unstable for $G_{SDC}=0.00$ nS. For small
values of $G_{SDC}$ bistability and hysteresis appears. The IP$_3$
range with hysteresis is largest for a $G_{SDC}$ value near 0.04 nS
(see Fig. \ref{fig5}). For higher values of $G_{SDC}$, the range of
hysteresis decreases until the typical Hopf-bifurcation for the
intracellular IP$_3$-mediated calcium oscillations remains for
$G_{SDC} = 0.20$ nS.

The IP$_3$ range of the hysteresis as a function on the SDC
conductance channel is shown in Fig. \ref{fig6}. We define the
IP$_3$ range of hysteresis as the range of [IP$_3$] in which
multiple states are found for increasing and decreasing [IP$_3$].
For example, in Fig. \ref{fig4} hysteresis takes place for [IP$_3$]
values between 0.45 and 1.75 $\mu M$, giving an IP$_3$ range of
hysteresis of 1.3 $\mu M$.

Recent data in the literature show that the SDC conductance, which
was found to give the largest range for hysteresis in our study
(near 0.04 nS), corresponds to the observed SDC conductance in other
studies \cite{parekh2005, krause1996, rychkov2005}. The SDC
conductance reported in Table 1 of \cite{parekh2005} and in
\cite{krause1996} was in the range between 0.04 and 0.05 nS (solid
line below the peak in Fig. \ref{fig6}). For hepatocytes
\cite{rychkov2005} a SDC conductance was reported in the range
between 0.08 and 0.14 nS. However, since the density of all ion
channels in hepatocytes is twice as high as in fibroblast
\cite{roosPhD,yin2005}, the ratio of conductances for the ion
channels is the same in hepatocytes and NRK fibroblasts. If we
correct for this higher density, rescaling all conductances for
those of NRK fibroblasts, 
we obtain the dotted line in \ref{fig6}. Therefore, the SDC
conductance, for which hysteresis is found over the largest range of
IP$_3$- values in our study (see Figs. \ref{fig5} and \ref{fig6}),
is in agreement with experimental observations for SDC conductance.

\section*{\small{Discussion}}

In this study we have analyzed a relatively simple model with an
excitable membrane and with IP$_3$-mediated calcium oscillations.
The interaction between these mechanisms in a single-cell model
revealed a surprisingly rich behavior with stable/instable states
with hysteresis.

The hysteresis and bistability of the membrane potential and the
intracellular calcium concentration in Fig. \ref{fig4}A and B
obtained by stability analysis of the single-cell model provide an
explanation for the various growth-state dependent changes in the
electrophysiological behavior of normal rat kidney (NRK) fibroblasts
in cell culture \cite{harks2005}. The stability analysis of the
single-cell model (Fig. \ref{fig4}) reveals for low [IP$_3$] values
(range 0.0 - 0.2 $\mu M$) a cell in the "resting state" . Increasing
the [IP$_3$] leads to spontaneous AP firing ("AP-firing state") and
at high [IP$_3$] values above 1.75 $\mu M$, the cell depolarizes
("depolarized state"). When we start at an [IP$_3$] value of 2.0
$\mu M$ and decrease [IP$_3$], the system is in the "depolarized
state" and changes from the "fast oscillating state" (range 0.45 -
0.8 $\mu M$) to the "AP-firing state"(range 0.2 - 0.45 $\mu M$) back
to its "resting state" (range 0.0 - 0.2 $\mu M$), which is in
agreement with experimental data shown by Harks et al.
\cite{harks2003b}. In the study of Harks et al. \cite{harks2005}
washout of the medium conditioned by the transformed cells by
perfusion with fresh serum-free medium, causes the cells to slowly
repolarize, and, preceded by a short period of fast small-amplitude
spiking of their membrane potential ("fast oscillating state"), to
regain spontaneous repetitive action potential firing activity
similar to that of the contact inhibited cells. This compares well
with the results in Fig. \ref{fig4}B, which shows for decreasing
[IP$_3$] a very similar behavior. Therefore, we conclude that the
stable states of the model as revealed by stability analysis of the
single-cell model correspond in great detail to the observed
growth-state dependent modulations of the membrane potential of NRK
fibroblasts in cell culture. This strongly suggests that these
growth-state dependent modulations of the membrane potential of NRK
cells reflect just different states of the same cell, rather than
the behavior of cells, that have differentiated to different cell
types with different properties during the various stages of
growth-factor stimulated development in vitro.

Most of the parameter values in our model were taken from the
literature (see \cite{kusters2005}) for a detailed overview).
Interestingly, the parameter values for the excitable membrane and
for the IP$_3$-mediated calcium oscillator, which are very different
mechanisms, are not independent. This can be understood from the
fact that the dynamics of the excitable membrane and of the IP$_3$
receptor are coupled by the cytosolic calcium concentration.
Changing one parameter of the excitable membrane or calcium
oscillator affects the other mechanism by changes in the cytosolic
calcium concentration. This is illustrated, for example, by Fig.
\ref{fig3}. Changing the strength of the SERCA pump causes large
differences in the range of hysteresis in the dynamics of cytosolic
calcium (Fig. \ref{fig3}), and therefore also in the dynamics of the
membrane potential (see Fig. \ref{fig4}, which illustrates the
relation between the dynamics of cytosolic calcium and the membrane
potential). Although the parameter values for the excitable membrane
and for the IP$_3$-mediated calcium oscillator were taken from
different studies, they fit nicely together to explain the behavior
of NRK cells both qualitatively and quantitatively. This provides
strong evidence for the reliability of these parameter values.
Moreover, this suggests that cells should have complicated
regulatory mechanisms to control all parameter values within a
proper range of parameter values to ensure the proper cell dynamics.

Summarizing, we explored the dynamical properties of a single-cell
model reproducing experimental observations on calcium oscillations
and action potential generation in NRK fibroblasts. A bifurcation
analysis revealed hysteresis and a complex spectrum of stable and
unstable states, which allows the system to switch among different
stable branches. Stability of the cell behavior is dominated by the
homeostatic function of the SDC channel. The conductance, which
provides the largest IP$_3$ range for hysteresis, compares well with
experimental values for this conductance \cite{parekh2005,
krause1996, rychkov2005}. Experimental observations
in NRK fibroblasts revealed the same kind of hysteresis as shown by this study. \\

We acknowledge financial support from the Nederlandse Organisatie
voor Wetenschappelijk Onderzoek (NWO), Ministerio de Educacion y
Ciencia (MEC), Junta de Andalucia (JA) and Engineering and Physical
Sciences Research Council (EPSRC), projects NWO 805.47.066, MEC
FIS2005-00791, JA FQM-165 and EPSRC EP/C0 10841/1.





\pagebreak

\begin{thebibliography}{10}

\bibitem{murrayBOOK}
Murray, J.~D.
\newblock (2002) {\em Mathematical {B}iology {I}. {A}n {I}ntroducion}.
\newblock (Springer, {N}ew {Y}ork).

\bibitem{keenerbook}
Keener, J.  \& Sneyd, J.
\newblock (1998) {\em Mathematical Physiology}.
\newblock (Springer, {N}ew {Y}ork).

\bibitem{haken1975}
Haken, H.
\newblock (1975) {\em Rev Mod Phys} {\bf 47}, 67--121.

\bibitem{jones1976}
Jones, B.~J.~T.
\newblock (1976) {\em Rev Mod Phys} {\bf 48}, 107--149.

\bibitem{velarde1977}
Normand, C., Pomeau, Y.  \& Velarde, M.~G.
\newblock (1977) {\em Rev Mod Phys} {\bf 49}, 581--624.

\bibitem{hohenberg1993}
Cross, M.~C.  \& Hohenberg, P.~C.
\newblock (1993) {\em Rev Mod Phys} {\bf 65}, 851--1112.

\bibitem{ashwin2005}
Ashwin, P.  \& Timme, M.
\newblock (2005) {\em Nature} {\bf 436}, 36--37.

\bibitem{huerta2001}
Rabinovich, M., Volkovskii, A., Lecanda, P., Huerta, R., Abarbanel, H.~D.  \&
  Laurent, G.
\newblock (2001) {\em Phys Rev Lett} {\bf 87}, 68--102.

\bibitem{laurent2001}
Laurent, G., Friedrich, M.~Stopfer R.~W., Rabinovich, M.~I., Volkovskii, A.  \&
  Abarbanel, H.~D.
\newblock (2001) {\em Annu Rev Neurosci} {\bf 24}, 263--297.

\bibitem{torres2002}
Pantic, L., Torres, J.~J., Kappen, H.~J.  \& Gielen, S.~C.
\newblock (2002) {\em Neural Comput} {\bf 14}, 2903--2923.

\bibitem{cortes2006}
Cortes, J.~M., Torres, J.~J., Marro, J., Garrido, P.~L.  \& Kappen, H.~J.
\newblock (2006) {\em Neural Comput} {\bf 18}, 614--633.

\bibitem{tsodyks2006}
Holcman, D.  \& Tsodyks, M.
\newblock (2006) {\em PLoS Comput Biol} {\bf 2}, e23.

\bibitem{sha2003}
Sha, W., Moore, J., Chen, K., Lassaletta, A.~D., Yi, C.~S., Tyson, J.~J.  \&
  Sible, J.C.
\newblock (2003) {\em Proc Natl Acad Sci USA} {\bf 100}, 975--980.

\bibitem{solomon2003}
Solomon, M.~J.
\newblock (2003) {\em Proc Natl Acad Sci USA} {\bf 100}, 771--772.

\bibitem{laurent1999}
Laurent, M.  \& Kellershohn, N.
\newblock (1999) {\em Trends Biochem Sci} {\bf 24}, 418--422.

\bibitem{ferrell2004}
Angeli, D., Ferrell, J.~E.  \& Sontag, E.~D.
\newblock (2004) {\em Proc Natl Acad Sci USA} {\bf 101}, 1822--1827.

\bibitem{ferrell2002}
Ferrell, J.E.
\newblock (2002) {\em Curr Opin Cell Biol} {\bf 14}, 140--148.

\bibitem{chay1985}
Chay, T.~R.  \& Rinzel, J.
\newblock (1985) {\em Biophys J} {\bf 47}, 357--366.

\bibitem{deyoung1992}
{De Young}, G.~W.  \& Keizer, J.
\newblock (1992) {\em Proc Natl Acad Sci USA} {\bf 89}, 9895--9899.

\bibitem{sneyd2002}
Sneyd, J.  \& Dufour, J.~F.
\newblock (2002) {\em Proc Natl Acad Sci USA} {\bf 99}, 2398--2403.

\bibitem{torres2004}
Torres, J.~J., Cornelisse, L.~N., Harks, E.~G.~A., {van Meerwijk}, W.~P.~M.,
  Theuvenet, A.~P.~R.,   \& Ypey, D.~L.
\newblock (2004) {\em Am J Physiol Cell Physiol} {\bf 287}, C851--C865.

\bibitem{kusters2005}
Kusters, J.~M. A.~M., Dernison, M.~M., {van Meerwijk}, W.~P.~M., Ypey, D.~L.,
  Theuvenet, A.~P.~R.  \& Gielen, C.~C. A.~M.
\newblock (2005) {\em Biophys J} {\bf 89}, 3741--3756.

\bibitem{harks2005}
Harks, E.~G.~A., Peters, P.~H.~J., {van Dongen}, J.~L.~J., {van Zoelen},
  E.~J.~J.  \& Theuvenet, A.~P.~R.
\newblock (2005) {\em Am J Physiol Cell Physiol} {\bf 289}, C130--C137.

\bibitem{ward2000}
Ward, S.M., \"{O}rd\"{o}g, T., Kohn, S.D., {Abu Baker}, S., Jun, J.Y., Amberg,
  G., Monaghan, K.  \& Sanders, K.M.
\newblock (2000) {\em J. Physiol.} {\bf 525}, 355--361.

\bibitem{dupont2003}
Dupont, G., Koukoui, O., Clair, C., Erneux, C., Swillens, S.  \& Combettes, L.
\newblock (2003) {\em FEBS Letters} {\bf 534}, 101--105.

\bibitem{harks2003}
Harks, E.~G., Torres, J.~J., Cornelisse, L.~N., Ypey, D.~L.  \& Theuvenet,
  A.~P.
\newblock (2003) {\em J Cell Physiol} {\bf 196}, 493--503.

\bibitem{harks2003b}
Harks, E.~G., Scheenen, W.~J., Peters, P.~H., {van Zoelen}, E.~J.  \&
  Theuvenet, A.~P.
\newblock (2003) {\em Pflugers Arch.} {\bf 447}, 78--86.

\bibitem{hofer1998}
Hofer, A.~M., Fasolato, C.  \& Pozzan, T.
\newblock (1998) {\em J. Cell Biol.} {\bf 140}, 325--334.

\bibitem{feske2006}
Feske, S., Gwack, Y., Prakriya, M., Srikanth, S., Puppel, S.H., Tanasa, B.,
  Hogan1, P.~G., Lewis, R.~S., Daly, M.  \& Rao, A.
\newblock (2006) {\em Nature} {\bf 441}, 179--185.

\bibitem{flourakis2006}
Flourakis, M., {Van Coppenolle}, F., Lehen$'$kyi, V., Beck, B., Skryma, R.  \&
  Prevarskaya, N.
\newblock (2006) {\em FASEB J.} {\bf 20(8)}, 1215--17.

\bibitem{deroos1997b}
{De Roos}, A.D., {Van Zoelen}, E.~J.  \& Theuvenet, A.~P.
\newblock (1997) {\em J Cell Physiol} {\bf 170}, 166--173.

\bibitem{fairgrave1991}
Fairgrieve, T.~F.  \& Jepson, A.~D.
\newblock (1991) {\em SIAM Journal on Numerical Analysis} {\bf 28}, 1446--1462.

\bibitem{ioossBOOK}
Iooss, G.  \& Joseph, D.
\newblock (1981) {\em Elementary {S}tability and {B}ifurcation {T}heory}.
\newblock (Springer, {N}ew {Y}ork).

\bibitem{ermentroutBOOK}
Ermentrout, B.
\newblock (2002) {\em Simulating, {A}nalyzing, and {A}nimating {D}ynamical
  {S}ystems. {A} {G}uide to {X}ppaut for {R}esearchers and {S}tudents}.
\newblock (SIAM, Philadelphia).

\bibitem{doedelAUTO}
Doedel, E.  \& Kernevez, J.~P.
\newblock (1986) {\em {AUTO}: {S}oftware for {C}ontinuation and {B}ifurcation
  {P}roblems in {O}rdinary {D}ifferential {E}quations}.
\newblock (California {I}nstitute of {T}echnology, {P}asadena, {CA}).

\bibitem{li2002}
Li, Y.  \& Rinzel, J.
\newblock (2002) {\em J Theor Biol} {\bf 166}, 461--473.

\bibitem{schuster2002}
Schuster, S., Marhl, M.  \& Hofer, T.
\newblock (2002) {\em Eur J Biochem} {\bf 269}, 1333--1355.

\bibitem{mignen2005}
Mignen, O., Brink, C., Enfissi, A., Nadkarni, A., Shuttleworth, T.~J.,
  Giovannucci, D.~R.  \& Capiod, T.
\newblock (2005) {\em J Cell Sci} {\bf 170}, 166--173.

\bibitem{parekh2005}
Parekh, A.~B.  \& Putney, J.~W.
\newblock (2005) {\em Physiol Rev} {\bf 85}, 757--810.

\bibitem{krause1996}
Krause, E., Pfeiffer, F., Schmid, A.  \& Schulz, I.
\newblock (1996) {\em J Biol Chem} {\bf 271}, 32523--32528.

\bibitem{rychkov2005}
Rychkov, G.~Y., Litjens, T., Roberts, M.~L.  \& Barritt, G.~J.
\newblock (2005) {\em Cell Calcium} {\bf 37}, 183--191.

\bibitem{roosPhD}
{de Roos}, A.~D.~G.
\newblock (1997) Ph.D. thesis (Katholieke Universiteit Nijmegen, The
  Netherlands).

\bibitem{yin2005}
Yin, Z.  \& Watsky, M.~A.
\newblock (2005) {\em Am J Physiol Lung Cell Mol Physiol} {\bf 288},
  L1110--L1116.

\end{thebibliography}

\pagebreak

\section*{Figure Legends}

\subsection*{Fig.~\ref{fig1}.}
Conceptual model of the processes involved in membrane excitability
and intracellular IP$_3$-mediated calcium oscillations by calcium
release through IP$_3$-receptors in the ER membrane in NRK
fibroblasts. Cell-membrane excitability is supported by inwardly
rectifying potassium channels (G$_{Kir}$), Ca-dependent Cl-channels
(G$_{Cl(Ca)}$), L-type Ca-channels (G$_{CaL}$), store-dependent
calcium (SDC) channels (G$_{SDC}$), a PMCA pump and leak channels
(G$_{lk}$). The total flux of calcium through the ER-membrane is the
result of the contribution by the SERCA pump, by the IP$_3$-receptor
($J_{IP_{3}R}$) and by leak channels in the $ER$ membrane
($J_{lkER}$). The membrane excitability and IP$_3$-mediated calcium
oscillations are coupled by the cytosolic calcium concentration,
which is also affected by a calcium buffer B.

\subsection*{Fig.~\ref{fig2}.}
Stable and unstable states for the excitable membrane using
$K_{lkER}$ as a control parameter. The intracellular Ca-oscillator
was silenced by setting the IP$_3$ concentration to zero. Thick
(thin) lines correspond to the stable steady-state solutions for the
[$Ca_{cyt}^{2+}$] (panel $A$) and for the membrane potential (panel
$B$) for increasing (decreasing) values of for $K_{lkER}$. The set
of parameter values in this model was as reported in
\cite{kusters2005}, with $G_{SDC} = G_{SOC}$ .

\subsection*{Fig.~\ref{fig3}.}
The bifurcation diagram for the intracellular calcium oscillator in
the single-cell model as a function of [IP$_3$] after elimination of
action potentials ($G_{CaL} = 0$) and with $J_{SERCA}^{ max}$ set to
2 x $10^{-5}$ $\mu mol/(s$ x $dm^2)$ in panel A and to 8 x $10^{-5}$
$\mu mol/(s$ x $dm^2)$ in panel B. Analogous to Fig. \ref{fig2},
thick and thin solid lines correspond to the stable states for
increasing and decreasing values for [IP$_{3}]$, respectively. The
three insets in panel A show the Ca-concentration as a function of
time for [IP$_3$] values at 0.01, 2 and 4 $\mu$M. The inset in B
shows the stable (solid lines) and unstable (dashed-dotted lines)
states for a large range of [IP$_3$] values. The set of all other
parameter values in this model was as reported in
\cite{kusters2005}, with $G_{SDC}= G_{SOC}$.

\subsection*{Fig.~\ref{fig4}.}

The bifurcation diagram for the single-cell model. The figure shows
the stable (solid lines) and unstable (dashed-dotted lines) states
for [$Ca_{cyt}^{2+}$] (panel $A$) and the membrane potential (panel
$B$) as a function of IP$_3$ concentration. Panel C and D show the
membrane potential for [IP$_3$] at 0.7 $\mu$M in case of increasing
and decreasing [IP$_3$], respectively. The small arrows on the
curves show the direction of change of the stable modes for
increasing and decreasing values of $[IP_3]$. The set of parameter
values in this model was as reported in \cite{kusters2005}.

\subsection*{Fig.~\ref{fig5}.}
Bifurcation diagrams for the single-cell model as shown in Fig.
\ref{fig4}A for different values of the $SDC$ channel conductance
from $G_{SDC}=0.02$ nS (bottom) to $G_{SDC}=0.20$ nS (top). At 0.04
nS (second graph from bottom), the hysteresis loop has a maximum in
the [IP$_3$] range from 0.5 to 1.95 $\mu$M.

\subsection*{Fig.~\ref{fig6}.}
\begin{figure}
\centerline{
\psfig{file=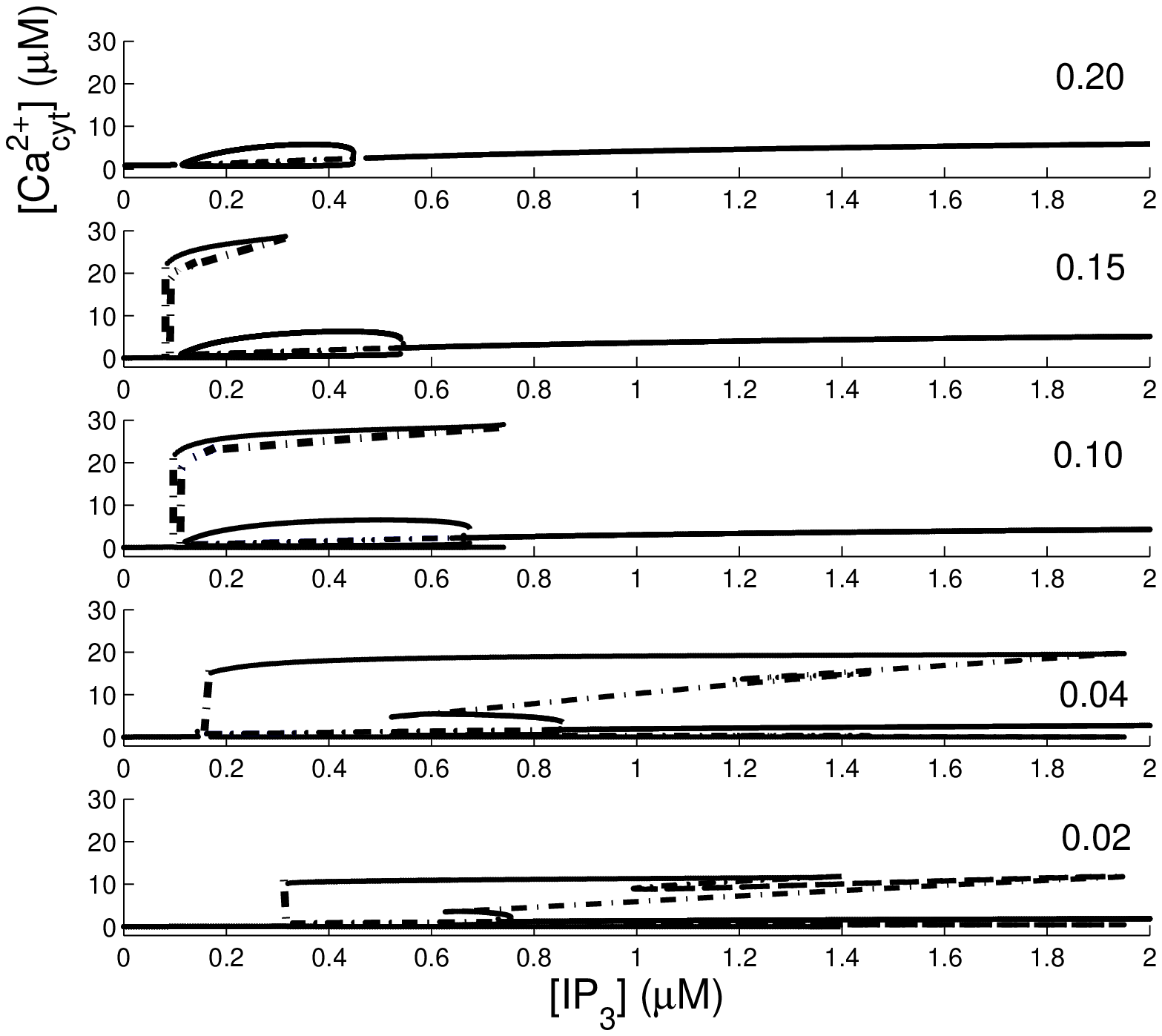,width=16cm}}\caption{}\label{fig5}
\end{figure}

\begin{figure}
\centerline{\psfig{file=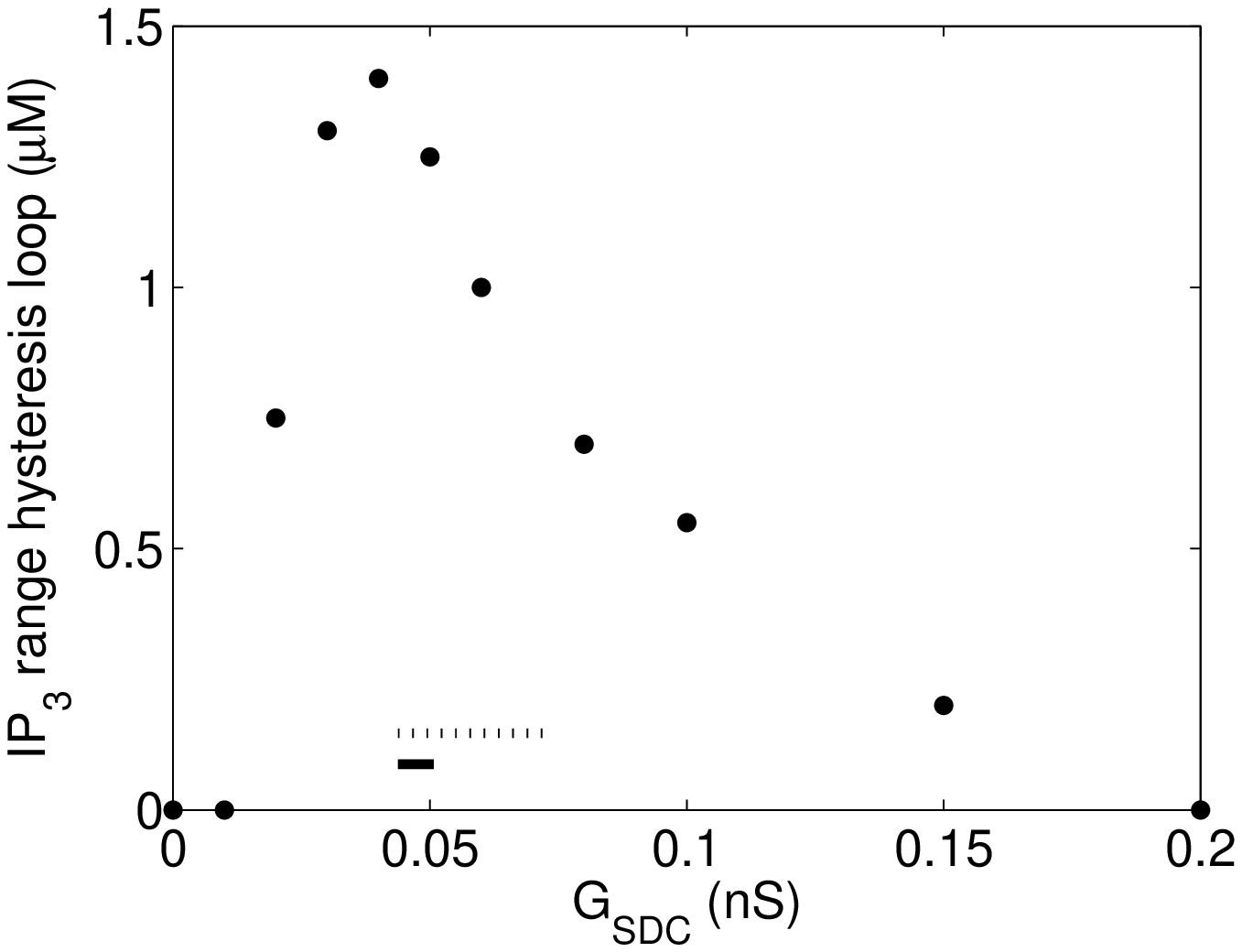,width=16cm}}\caption{}\label{fig6}
\end{figure}

IP$_3$ range of the hysteresis loop in $\mu$M (cf. Fig. \ref{fig5})
as a function on SDC conductance ($G_{SDC}$) for a NRK cell with a
capacitance of 20 pF reveals a value in which the hysteresis area is
maximum (dots are simulated data). Below the peak, we illustrate the
$SDC$ conductance measured in experiments and separately reported in
\cite{parekh2005,krause1996} (solid line) and \cite{rychkov2005}
(dashed line). \pagebreak

\end{document}